# An Extra Electrostatic Energy in Semiconductors and its Impact in Nanostructures

Jean-Michel Sallese

Ecole Polytechnique Fédérale de Lausanne - Switzerland

PACS: 41.20.Cv

**ABSTRACT.**

This work revisits the classical concept of electric energy and suggests that the common definition is likely to generate large errors when dealing with nanostructures. For instance, deriving the electrostatic energy in semiconductors using the traditional formula fails at giving the correct electrostatic force between capacitor plates and reveals the existence of an extra contribution to the standard electrostatic energy. This additional energy is found to proceed from the generation of space charge regions which are predicted when combining electrostatics laws with semiconductor statistics, such as for accumulation and inversion layers. On the contrary, no such energy exists when relying on electrostatics only, as for instance when adopting the so-called full depletion approximation. The same holds for charged or neutral insulators that are still consistent with the customary definition, but which are in fact singular cases. In semiconductors, this additional free energy can largely exceed the energy gained by the dipoles, thus becoming the dominant term. Consequently, erroneous electrostatic forces in nanostructure systems such as for MEMS and NEMS as well as incorrect energy calculations are expected using the standard definition. This unexpected result clearly asks for a generalization of electrostatic energy in matter in order to reconcile basic concepts and to prevent flawed force evaluation in nanostructures with electrical charges.

## I    Introduction.

Interpretation of electric energy in conjunction with thermodynamics has been widely investigated, with a special interest for dielectric bodies and ideal conductors [1-5]. The electric energy stored inside of a body can be expressed whether in terms of charges and potentials restricted to the volume of the body, or in terms of fields including contributions beyond the physical boundary of the system, see relations (1) and (2). Careful considerations and exhaustive criticisms about the validity of electrostatic energy formulation in conductors and insulators have been addressed in [1, 2]. Beyond the thorough literature, instructive and complete analysis of thermodynamics of electric and magnetic fields has also been developed in [4, 6].

While the topic of electric energy in matter seems to be well established and widely accepted, we will bring some evidence that this paradigm must be revisited in semiconductors for instance, resulting in a new contribution to the well-known Helmholtz free energy arising from electric fields.

Before discussing an instructive virtual experiment, we revisit some fundamental relationships for electrostatic energy from a thermodynamics point of view.

The electrical work $U_e$ that must be spent to gather charges from infinity into a volume $\Omega$ is given by [3,5,6]:

$$U_e = \frac{1}{2} \int_\Omega \rho \, \psi \, d\Omega \tag{1}$$

where $\rho$ is the local charge density and $\psi$ is the electric potential (this expression includes the self energy contribution [5]). The integral is limited to the volume of the body containing the charges, and in this sense, relation (1) represents the electric energy of the content of $\Omega$, which is also part of the internal energy. Then, the internal energy of electrical nature is expected to be implicitly contained in (1), which is indeed how Frankl [7] analyzes the free energy stored in the depletion regions at the surface of a silicon layer. We will come to that point later.

This formulation attributes energy to electric charges. Alternatively, adopting the electric field and displacement vectors concepts ($\boldsymbol{E} = -grad(\psi)$, $div(\boldsymbol{D}) = \rho$; bold letters hold for vectors), the electrostatic energy can also be expressed from the electric field generated by the charges enclosed in the volume $\Omega$ provided the integration is performed over the whole space $\Omega_\infty$, *including matter:*

$$U_e = \frac{1}{2} \int_{\Omega_\infty} \boldsymbol{E} \cdot \boldsymbol{D} \, d\Omega \tag{2}$$

The electric field in (2) does not account for all sources in the universe; it is only assigned to the charges located in the volume $\Omega$. It can be shown that equations (1) and (2) are equivalent and represent the electrostatic energy of those charges.

A generalisation of relation (2) to nonlinear polarisable materials [1-5] is given by:

$$U_e = \int_{\Omega_\infty} \left( \int_0^{\boldsymbol{D}} \boldsymbol{E} \cdot d\boldsymbol{D} \right) d\Omega \tag{3}$$

Relation (3) is a priori very general and relies on the incremental work spent upon creating the electric field in matter and in free space, without any assumption on the relation linking $\boldsymbol{E}$ to $\boldsymbol{D}$. From a thermodynamic point of view, this represents the Helmholtz free electric energy of the system when assuming an isothermal process under constant deformation [1, 3].

The energy belongings to dipoles [1,2] is also included in (3) through the polarization vector $\boldsymbol{P}$ that satisfies $\boldsymbol{D} = \varepsilon_0 \cdot \boldsymbol{E} + \boldsymbol{P}$. Finally, the electrostatic energy is made up of two contributions, i.e. $U_e = U_f + U_P$, where $U_f$ is the electric field energy, valid in matter as in free space:

$$U_f = \frac{1}{2}\varepsilon_0 \int_{\Omega_\infty} |\boldsymbol{E}|^2 \cdot d\Omega \qquad (4)$$

whereas $U_P$ is an energy related to polarization processes experienced by the body $\Omega$ [2]:

$$U_P = \int_\Omega \left( \int_0^{\boldsymbol{P}} \boldsymbol{E} \cdot d\boldsymbol{P} \right) d\Omega \qquad (5)$$

In dielectrics, $U_P$ can be thought as a transformation of electric energy in some internal energy that belongs to the body. Obviously, this term cancels in ideal conductors since no electric field penetrates in the volume. However, concerning semiconductors, we can wonder if $U_P$ is still the only contribution to the Helmholtz free energy of electric nature. In this work, we propose to analyze how electric energy is stored in semiconductors and if this obeys to the same law as for dielectrics. To the best of our knowledge, a detailed transfer of electric energy in semiconductors has never been examined so far.

## II  Virtual experiment with semiconductor based capacitors.

In this section, we analyze the work spent upon moving a semiconductor based capacitor plate with respect to a counter conductor plate, and compare it with the variation of the Helmholtz free energy predicted from relations (1) or (2).

For this purpose, we will consider two semi-infinite capacitors, *C0* and *C1*, with plates connected such as in figure 1. Except for *C1* where one electrode is a semiconductor (p-type doped, without loss of generality), other electrodes are made of ideal conductors. This special arrangement will simplify the thermodynamic analysis as there is no need for introducing any voltage source. The case of capacitors biased with an external voltage source has been inspected in many details by Bobbio [1, 2]. This will also be introduced in the last section to generalize the analysis for arbitrary geometries.

### 1) Derivation of the total electrostatic energy.

Applying the definition of the electric energy given by relation (2) and assuming that electric charges on opposite electrodes compensate each other (we consider semi-infinite plates, i.e. the electric field in the free space surrounding the capacitors is null) and using scalars instead of vector (consistent with the axis orientation), we obtain:

$$U_0 = \int_{\Omega_\infty} \left( \int_0^{D_0(x)} E\, dD \right) dx = \frac{1}{2}\varepsilon_0 E_{g0}^2 g_0 \qquad (6)$$

$$U_1 = \int_{\Omega_\infty} \left( \int_0^{D_1(x)} E\, dD \right) dx = \frac{1}{2}\varepsilon_0 E_{g1}^2 g_1 + \int_{-d}^0 \left( \int_0^{D_1(x)} E\, dD \right) dx \qquad (7)$$

Here $\Omega_\infty$ represents the whole space (including volume between capacitor plates), $g_0$ and $g_1$ are the electrodes gap for *C0* and *C1* respectively, $E_{g0}$ and $E_{g1}$ are the *uniform* scalar electric fields between electrodes (according to figure 1 $E_{g0}, E_{g1} \geq 0$) and *d* is the semiconductor thickness. In addition, since for linear materials (semiconductors) we have $D = \varepsilon_{sc} E$ ($\varepsilon_{sc}$ is the semiconductor permittivity), relation (7) becomes:

$$U_1 = \frac{\varepsilon_0}{2} E_{g1}^2 g_1 + \frac{\varepsilon_{sc}}{2} \int_{-d}^0 E^2\, dx = \frac{\varepsilon_0}{2} E_{g1}^2 g_1 - \frac{\varepsilon_{sc}}{2} \int_{-d}^0 E \cdot \frac{d\psi}{dx} \cdot dx = \frac{\varepsilon_0}{2} E_{g1}^2 g_1 - \frac{\varepsilon_{sc}}{2} \int_0^{\psi_S} E \cdot d\psi \qquad (8)$$

where $\psi_s = -\int_{-d}^0 E\, dx$ is the potential drop between the surface of the semiconductor and the neutral body, i.e. the surface potential [8] ($\psi_S \leq 0$, $\psi(0)=\psi_S$, $\psi(-d)=0$ according to figure 1). A description in terms of conduction and valence bands is shown on figure 2 where, for convenience, the higher potential was applied to the P type semiconductor.

From relation (2), the electrostatic energy of the global system is :

$$U_e = U_0 + U_1 = \frac{\varepsilon_0}{2} E_{g0}^2 g_0 + \frac{\varepsilon_0}{2} E_{g1}^2 g_1 - \frac{\varepsilon_{sc}}{2} \int_0^{\psi_S} E \cdot d\psi \qquad (9)$$

### 2) Electric work upon displacement: need for a new energy term.

In this virtual experiment, tied electrodes are pre-charged with a total average charge density (per unit surface) $Q_T$, before being isolated, see figure 1. Any displacement *dx* of the semiconductor counter electrode (others are supposed fixed) will induce a variation in the electrostatic energy for *C0* and *C1*; as well as a mechanical work $\delta W_F$ arising from the attractive electrostatic force *F* between the plates.

Invoking the fundamental law of thermodynamics, when the displacement is performed at constant total charge, *i.e.* no connection to any kind of voltage source, we can write (assuming that all bodies have the same temperature):

$$\delta W_F = -\mathbf{F} \cdot d\mathbf{x} = dU_f + \sum_{conductors} dU_C + \sum_{semiconductors} dU_{SC} - \sum_{conductors} T \cdot dS_C - T \cdot dS_{SC} \qquad (10)$$

As in [2], we introduce the Helmholtz free energy for each of the bodies. Relation (10) becomes:

$$\delta W_F = dU_f + \sum_{conductors} dA_C + \sum_{conductors} S_C \cdot dT_C + dA_{SC} + S_{SC} \cdot dT_{SC} \tag{11}$$

where $dU_f$ is the variation of the electric field energy in the whole space (including bodies), $dA_C$ and $dA_{SC}$ are the changes of the Helmholtz free energy of the capacitor plates induced by a displacement $dx$ while maintaining constant the total charge $Q_T$, and $S_c$, $S_{sc}$ represent their respective entropy. Note that for dielectrics, the Helmholtz free energy reverts to $U_P$ [2].

As in [2], we assume that the temperature of the system is maintained constant. The displacement having been performed at constant free charges, no other work, except that induced by the displacement of the semiconductor electrode is spent upon the system. Assuming that there is no deformation [2], $dA_c=0$, relation (11) simplifies into:

$$\delta W_F = dU_f + dA_{SC} \tag{12}$$

Bobbio [2] demonstrated that the term in the right hand side of (12) is nothing but the electric energy as defined from (4), *i.e.* $\delta W_F = dU_e$.

The way $Q_T$ will redistribute among (connected) electrodes will depend on the total charge density and the electrodes gaps $g_0$ and $g_1$, as well as on the physical nature of the plates, *i.e.* metals or semiconductors. In addition, given that connected plates must share the same potential, we have ( $E_{g0}, E_{g1} \geq 0$; $\psi_S \leq 0$ ):

$$E_{g0}\, g_0 = E_{g1}\, g_1 - \psi_S \tag{13}$$

Since the overall charge density $Q_T$ is fixed, Gauss theorem imposes that the sum of the electric fields $E_{g1}$ and $E_{g0}$ is invariant:

$$\delta(E_{g0} + E_{g1}) = 0 \tag{14}$$

Next, from the continuity of the displacement vector, without presuming for any charge sheet on the semiconductor surface, the electric fields across the semiconductor/air interface satisfy:

$$\varepsilon_{sc} E_s = \varepsilon_0 E_{g1} \tag{15}$$

where $E_S$ is the surface electric field evaluated *inside* the semiconductor.

Differentiating (9) gives the variation of the total electric energy for the capacitors system:

$$dU_e = \left(\varepsilon_0\, E_{g0}\, \partial E_{g0}\, g_0\right) + \left(\varepsilon_0\, E_{g1}\, \partial E_{g1}\, g_1 + \frac{\varepsilon_0}{2} E_{g1}^2 \cdot \partial g_1\right) - \frac{\varepsilon_{sc}}{2} \partial \int_0^{\psi_S} E \cdot d\psi \tag{16}$$

Merging relations (13), (14) with (16), the change in electric energy can be expressed in terms of *C1* variables only:

$$dU_e = \frac{\varepsilon_0}{2} E_{g1}^2 \cdot \partial g_1 + \partial E_{g1} \cdot \varepsilon_0\, \psi_S - \frac{\varepsilon_0}{2} E_{g1} \cdot \partial \psi_S \tag{17}$$

Finally, using (15) in (17), the force acting on the semiconductor plate for *C1* as derived from the electric energy becomes:

$$F_U = \frac{dU_e}{dg_1} = \frac{\varepsilon_0}{2} E_{g1}^2 + \frac{\partial E_{g1}}{\partial g_1} \cdot \varepsilon_0 \ \psi_S - \frac{\varepsilon_0}{2} E_{g1} \cdot \frac{\partial \psi_S}{\partial g_1} = \frac{\varepsilon_0}{2} E_{g1}^2 + \varepsilon_{sc} \frac{\partial E_S}{\partial g_1} \cdot \psi_S - \frac{\varepsilon_{sc}}{2} E_S \cdot \frac{\partial \psi_S}{\partial g_1} \qquad (18)$$

We call this force the 'field energy force' since it is derived from electric field and dipoles energies, i.e. relation (3) (the sign is consistent with the axis orientation).

On the other hand, the electric charge density on *C1* will also create an attractive force between the plates which is known as the Coulomb force, noted $F_C$. This force is the product of the total charge in the semiconductor plate times the electric field generated by the counter electrode (which is half the electric field in the gap), excluding the contribution of the charge itself ($F_c$ is positive according to figure 1):

$$F_C = \left(\varepsilon_0 E_{g1}\right) \cdot \frac{1}{2} E_{g1} = \frac{1}{2} \varepsilon_0 E_{g1}^2 \qquad (19)$$

Relation (19) should then be regarded as the 'true' force. As such, the force derived from the free energy (relation 18) and the Coulomb force should be strictly equivalent.

However, relations (18) and (19) are not equal.

The inconsistency between these two formula is a major result *per se* and suggests that the definition of the electric energy given by (3) does not represent the total electrostatic energy gained by the system.

Without loss of generality, a new contribution of electric nature to the Helmholtz free energy, that we call $A_{Extra}$, needs to be introduced. Adopting this term, relation (12) becomes:

$$\partial W_F = dU_e + dA_{Extra} = dU_f + dA_{SC} + dA_{Extra} \qquad (20)$$

Now, imposing $F_C$ and $F_U$ to be equal, this extra energy must satisfy:

$$\frac{dA_{Extra}}{dg_1} + \varepsilon_{sc} \frac{\partial E_S}{\partial g_1} \cdot \psi_S - \frac{\varepsilon_{sc}}{2} E_S \cdot \frac{\partial \psi_S}{\partial g_1} = 0 \qquad (21)$$

Next, integrating from $-\infty$ to $g_1$, and noting that the electric field and the surface potential must vanish at infinity, we obtain:

$$A_{Extra} = \frac{\varepsilon_{sc}}{2} \int_0^{\psi_S} E_S \cdot d\psi - \varepsilon_{sc} \cdot \int_0^{E_S} \psi_S \cdot dE = \varepsilon_{sc} \cdot \left( \frac{3}{2} \int_0^{\psi_S} E \cdot d\psi - \psi_S \cdot E_S \right) \qquad (22)$$

This result is essential. It proves that in addition to the free energy $U_p$ belonging to the dipoles, a new free energy notion is predicted. Theoretically, this is a central concept which was not anticipate if we concede that relation (2) was hold to account for the energy of electric nature in matter and in free space.

We can go one step further. Introducing the electric susceptibility $\chi = \varepsilon_{sc}/\varepsilon_0 - 1$ and assuming the medium to be linear, the polarization vector can be written $\boldsymbol{P} = \varepsilon_0 \cdot \chi \cdot \boldsymbol{E}$ and the free energy arising from the polarization becomes:

$$U_P = \frac{-\varepsilon_0 \cdot \chi}{2} \cdot \int_0^{\psi_S} E \cdot d\psi \tag{23}$$

Combining (23) with (22) finally gives:

$$A_{Extra} = -\varepsilon_{sc} \cdot \left( \psi_S \cdot E_S + \frac{3}{\varepsilon_0 \cdot \chi} U_P \right) \tag{24a}$$

Therefore, in a linear medium, the additional free energy is a linear function of the dipole free energy with the product of the 'internal' potential drop inside the body itself with the electric field at the free surface.

The total free energy belonging to the semiconductor becomes:

$$U_{SC} = A_{Extra} - \frac{\varepsilon_{sc}}{2} \int_0^{\psi_S} E_S \cdot d\psi = \varepsilon_{sc} \cdot \int_0^{E_S} \psi \cdot dE \tag{24b}$$

which is the upper surface in the $E$-$\Psi$ plot of figure 3. This sound like a co-energy representation.

### III) The case of neutral and charged insulators.

There are situations where the electrostatic force rebuilt from relations (4) and (5) works correctly. We will provide some evidence that this does happen for insulators, and more generally for charged insulators, which are *de facto* traditional systems coming along with electric energy considerations.

We replace the semiconductor by an insulator in the system of coupled capacitors sketched in figure 1. Next, we will calculate the Coulomb $F_C$ and energy based $F_U$ forces and assess if these are equal.

Regarding the energy stored in the capacitor *C1*, relation (7) can be rewritten in terms of the electric field in the insulator $E_i$:

$$U_1 = \frac{\varepsilon_i}{2} \cdot \int_{-d}^{0} E_i^2(x) \, dx + \frac{1}{2} \varepsilon_0 \, E_{g1}^2 \, g_1 \tag{25}$$

*($\varepsilon_i$ is the insulator dielectric constant).*

Introducing the local charge density in the insulator $\rho(x)$ and assuming that $\rho(x)$ does not depend on the local potential, the integration of the Poisson equation in the insulating layer gives:

$$E_i(x) = \frac{1}{\varepsilon_i} \cdot \int_0^x \rho(u) \cdot du + E_i(0) \tag{26}$$

Since the charge density only depends on the coordinate, we can express the electric field in the form:

$$E_i(x) = f(x) + E_i(0) \tag{27}$$

where $f(x)$ is a function of the coordinate only.

Similarly, integration of (27) gives the potential distribution in the insulator:

$$\psi(x) = \psi(0) - \int_0^x f(u) \cdot du - x \cdot E_i(0) \tag{28}$$

After manipulation, the total electric energy $U_e = U_1 + U_0$ can be written as:

$$U_e = \frac{\varepsilon_i}{2} \cdot \left[ \int_{-d}^0 f^2(u) \cdot du + 2 \cdot E_i(0) \cdot \int_{-d}^0 f(u) \cdot du + E_i^2(0) \cdot d \right] + \frac{\varepsilon_0}{2} E_{g1}^2 \, g_1 + \frac{\varepsilon_0}{2} E_{g0}^2 \, g_0 \tag{29}$$

Noting that integrals in the bracket of (29) do not depend on the local potential, and thus on the value of the gap $g_1$, we can calculate the force arising from the expression of the electric energy:

$$F_U = \frac{dU_e}{dg_1} =$$
$$\varepsilon_i \cdot \frac{dE_i(0)}{dg_1} \cdot \int_{-d}^0 f(u) \cdot du + \varepsilon_i \cdot d \cdot \frac{dE_i(0)}{dg_1} E_i(0) + \frac{\varepsilon_0}{2} E_{g1}^2 + \varepsilon_0 E_{g1} \frac{dE_{g1}}{dg_1} g_1 + \varepsilon_0 E_{g0} \frac{dE_{g0}}{dg_1} g_0 \tag{30}$$

Further, the potential on connected plates must be equal. Using (28), this reads:

$$E_0 \cdot g_0 = E_1 \cdot g_1 + \psi(-d) - \psi(0) = E_1 \cdot g_1 - \int_0^{-d} f(u) \cdot du + d \cdot E_i(0) \tag{31}$$

Relation (30) becomes:

$$\frac{dU_e}{dg_1} = \varepsilon_i \cdot \frac{dE_i(0)}{dg_1} \cdot \int_{-d}^0 f(u) \cdot du + \varepsilon_i \cdot \frac{dE_i(0)}{dg_1} E_i(0) \cdot d + \frac{\varepsilon_0}{2} E_{g1}^2 + \varepsilon_0 E_{g1} \frac{dE_{g1}}{dg_1} g_1$$
$$+ \varepsilon_0 \frac{dE_{g0}}{dg_1} \left( E_1 \cdot g_1 - \int_0^{-d} f(u) \cdot du + d \cdot E_i(0) \right) \tag{32}$$

Again, charge conservation writes:

$$dE_{g1} = -dE_{g0} \tag{33}$$

Similarly, continuity of the displacement vector at the dielectric-air interface gives:

$$\varepsilon_i E_i(0) = \varepsilon_0 E_{g1} \tag{34}$$

Finally, after simplifications, we find that the electrostatic force based on the electric energy variation upon electrode displacement reverts to the Coulomb force between (infinite) charged capacitor plates:

$$\frac{dU_e}{dg_1} = F_U = \frac{1}{2}\varepsilon_0 E_{g1}^2 \tag{35}$$

We can conclude that as far as insulators are concerned, there is no need to introduce any new free energy of electric nature. This apparent 'trivial' result for insulators could explain why the inconsistency that will be raised for semiconductors has been concealed.

### IV)   Sufficient condition for the existence of the extra free energy.

In the quest for a more general criteria, we can analyze when the extra energy given by (22) vanishes, i.e. when the electric energy in a system is still given by (3). Following former analysis, this condition is verified as soon as $F_U=F_C$ in relations (18) and (19). Imposing this identity links the surface electric field to the surface potential:

$$\frac{dE_S}{E_S} = \frac{1}{2}\frac{d\psi_S}{\psi_S} \tag{36}$$

which trivial solutions are:

$$E_S = C \cdot \sqrt{-\psi_S} \quad \text{when} \quad \psi_S \leq 0 \qquad \text{or} \tag{37a}$$

$$E_S = -C \cdot \sqrt{\psi_S} \quad \text{when} \quad \psi_S \geq 0 \tag{37b}$$

Where $C$ is a positive valued integration constant.

A special and highly interesting case occurs when this condition is taken valid inside the whole volume of the body, and not only at its surface. In that case when we can write:

$$\frac{dE(x)}{E(x)} = \frac{1}{2}\frac{d\psi(x)}{\psi(x)} \tag{38}$$

Without loss of generality, we assume $\psi_S \leq 0$ as in figure (2). Again, the solution is:

$$E(x) = C \cdot \sqrt{-\psi(x)} \tag{39}$$

Making use of the Poisson equation, we find that the charge density must be constant (and negative in our case) where the electric field exists:

$$\frac{dE(x)}{dx} = \frac{-\rho(x)}{\varepsilon_I} = \frac{C^2}{2} \tag{40}$$

(similarly when $\psi_S \geq 0$ the same conclusion applies with a positive charge).

Basically, for semiconductors this condition reverts to the so-called full depletion approximation in a uniformly doped material [8]. It imposes that the body is expected to be fully depleted down to a given coordinate until it changes for neutrality in a step-like transition. This analysis only makes use of the Poisson equation and is a quite common approximation in doped semiconductors biased in depletion..

However, as it will be discussed in the next section, adopting a modeling approach involving Fermi-Dirac or Boltzmann statistics rules out this quite crude depletion approximation and will introduce a new energy term. Therefore, we can state that the additional energy gained by the semiconductor should root in statistical physics.

## V) The special case of semiconductors.

In order to estimate the magnitude and impact of the free energy given by (22), semiconductors represent a class of ideal systems since relatively simple analytical expressions link the surface potential to the surface electric field. Considering a non-degenerate p-type doped silicon layer, the surface potential and the surface electric field are related through the well-known relationship, valid for depletion, inversion and accumulation [8] (signs are consistent with figure 2):

$$E_S = -sign(\psi_s)\sqrt{\frac{2 U_T q N_A}{\varepsilon_{sc}}} \sqrt{\frac{n_i^2}{N_A^2}\left(e^{\frac{\psi_S}{U_T}} - \frac{\psi_S}{U_T} - 1\right) + \left(e^{\frac{-\psi_S}{U_T}} + \frac{\psi_S}{U_T} - 1\right)} \qquad (41)$$

where $U_T$ is the thermal voltage, $n_i$ the intrinsic carrier density, $N_A$ the p-type doping concentration, other symbols having their usual meaning (note that (41) is also valid at each coordinate inside the semiconductor). Adopting the conventional representation $E_S(\psi_S)$ and considering only positive values of the surface potential (depletion-inversion modes), the main contributions to $A_{extra}$ as defined by the two integrals in (22) which correspond to adjacent surfaces in the $E_S$-$\Psi_S$ plot in figure 3 (doping is $10^{16}$ cm$^{-3}$).

This quite intuitive interpretation of $A_{extra}$ is interesting: for a doping of $10^{16}$cm$^{-3}$, increasing the surface potential beyond what is called *'onset of inversion'* (see figure 3), which corresponds approximately to $\Psi_S$ =0.7 volt in our case, will tend to increase exponentially the weight of the 'upper' surface with respect to the 'lower' one, and so it will also exponentially increase the amount of free energy $A_{extra}$. The same holds for negative value of $\Psi_S$, i.e. accumulation mode, where the dominance of $A_{extra}$ is even more striking.

While $\Psi_S$ remains below 0.7 volt (doping $10^{16}$cm$^{-3}$), relation (41) can be fairly well approximated by:

$$E_S = -\sqrt{2\ q N_A/\varepsilon_{sc}} \cdot \sqrt{\psi_S} \qquad (42)$$

Except for the sign of $\Psi_S$, relation (42) is formally the same as (37b) and so $A_{extra}$ is expected to be negligible. Therefore, we anticipate that as far as the surface potential is above 0.7 volt or possibly negative, $A_{extra}$ will contribute substantially to the total electric free energy.

In order to compare how the Helmholtz free electric energy is shared between $A_{Extra}$ and $U_P$ (see relation 5), we propose to evaluate the ratio $A_{Extra}/U_P$ versus the surface potential for various silicon doping densities.

As foreseen, figure (4a) reveals that the electric energy subsequent to a change in the semiconductor charge density ($Q_{SC}$) becomes dominant as soon as we enter in accumulation, i.e. $\psi_S < 0$, or in strong inversion (*i.e.* $\psi_S > 0.7V$ for $N_A=10^{16}cm^{-3}$). This inversion limit is shifted towards lower surface potentials when the doping density is decreased, whereas for accumulation the doping density has almost no effect when using such a surface potential representation.

The same information, but now in terms of charges, is illustrated on figure 4b. It reveals that $A_{Extra}$ always dominates $U_P$ in accumulation, whereas for depletion-inversion mode there exist some ranges where $A_{extra}$ is small with respect to $U_P$. Still, this has to be relate to the full depletion assumption which is partly verified before strong inversion takes place. Whereas the doping density had no visible effect in accumulation, here these tiny changes are magnified by the exponential link between the charge density and the surface potential.

Interestingly, the asymptotic case where the semiconductor is neutral ($\psi_S \approx 0V$) gives $A_{Extra}/U_P \approx 3\varepsilon_{SC}/\varepsilon_0 \cdot \chi$, which is about *1.09*. It is worth noticing that this ratio is independent of the doping.

**Analysis of electrostatic forces derived from the 'common' free energy formula.**

For simplicity, electrode gaps $g_1$ and $g_0$ have been given the same value. Two gap separations have been used, namely 100 and 10 nm, as well as different doping densities. Regarding the total charge density (per unit surface) $Q_T$, it has been assigned values consistent with MOS and MEMS based devices.

Solving the set of equations involving basics of electrostatics and semiconductors gives the charge densities on each of the plates, as well as the surface potential for the semiconductor. Next, the ratio $F_C/F_U$ has been computed and displayed on figures 5a and 5b for the values of $g_{1,0}$.

When the distance between electrodes is set to 100 nm, figure (5a) confirms that $F_C$ and $F_U$ given by relation (18) are indeed not equal, even though this difference is not so high for substrate doping densities greater than $10^{16}$ cm$^{-3}$. However, the discrepancy is increased when the doping is lowered down to $10^{15}$ cm$^{-3}$.

Decreasing the gap down to 10 nm is even more instructive. Figure (5b) reveals that $F_C$ and $F_U$ start to differ almost by one order of magnitude for a doping density of $10^{15}$ cm$^{-3}$, and by a factor of about 4 even for highly doped cases.

This observation can be understood as follows: as far as the energy stored in free space exceeds the energy stored inside of the body, the difference in $F_U$ and $F_C$ will remain small. This happens for relatively 'large' systems with micrometer based gap electrodes. But when electric energies stored inside and outside the semiconductor become comparable, those forces will start to deviate from one another significantly. As rule of thumb, we can say that $A_{Extra}$ cannot be neglected as soon as the gap between the plates compares with the extension of non-neutral regions in the semiconductor. This value ranges typically from sub-micrometers in low-doped semiconductors, down to dozens of nanometers in highly doped ones.

But no matter the magnitude of these effects: here it is crucial to note that these striking results follow straight from the bare application of the definition of electric energy as given from relation 3, at least when semiconductors are of interest.

## VI) Generalisation to arbitrary geometries.

The simple picture of a 1D ideal semi-infinite capacitor system revealed the presence of a new free energy in semiconductors. Generalization to three dimensional systems is still an issue and needs further developments. In this section, we propose to generalize the analysis by considering a three dimensional body, *i.e.* a semiconductor 'SC' in our case, which is separated from an ideal conductor 'M' by vacuum, as shown on figure 6.

The virtual experiment by which the electric energy is transferred to the semiconductor body will now proceed as follow: the body is maintained fixed and no mechanical work is done, i.e. $\delta W_F = 0$. However, the potential of a voltage source $V_P$ is continuously increased (quasi-static charging) until a given bias, whereas the potential of the conductor plate $V_C$ is maintained at 0 volt. The $V_P$ potential is applied at the locus $P$ in the semiconductor.

As in section II, we assume a rigid body maintained at a fixed temperature. Then, the energy gained by the system {semiconductor, conductor} reverts to the energy spent by the voltage source ($V_{cp}=V_c - V_p$), we can write:

$$\delta W_{vs} = -\delta W_{system} = V_{CP}(Q_{sc})\delta Q_{sc} \qquad (43)$$

where $Q_{sc}$ represents the total charge in the semiconductor.

According to Gauss theorem, a variation of the semiconductor charge reverts to a variation of the flux of the surface displacement vector $d\mathbf{D}$ through the surface surrounding the body $S_1$:

$$\delta W_{vs} = V_{CP}(Q_{sc}) \oint_{S_1} d\mathbf{D} \cdot d\mathbf{S} \qquad (44)$$

where $d\mathbf{S}$ is the unitary vector normal to the surface (oriented outwards).

Since the potential $V$ is a constant, we can write:

$$\delta W_{vs} = \oint_{S_1} V_{CP}(Q_{sc})(d\mathbf{D} \cdot d\mathbf{S}) \qquad (45)$$

Figure 6 shows a line $\mathbf{L}$ joining the point $P$ on the semiconductor surface to the conductor $C$. This line crosses the surface $S$ of the semiconductor at $N$. Since the potential drop between $P$ and $C$ does not depend on the path (the voltage drop is zero in a closed loop), we can write (the dependence on $Q_{sc}$ is implicit for potentials):

$$\delta W_{vs} = \oint_{S_1} [(V_C - V_N) + (V_N - V_P)](d\boldsymbol{D} \cdot d\boldsymbol{S}) \qquad (46)$$

In relation (46), $(V_N - V_P)$ represents the potential drop in the semiconductor at point $N$, and this reverts to the surface potential $\psi_S$ in our former discussion:

$$\delta W_{vs} = \oint_{S_1} \psi_S (d\boldsymbol{D} \cdot d\boldsymbol{S}) + \oint_{S_1} (V_C - V_N)(d\boldsymbol{D} \cdot d\boldsymbol{S}) \qquad (47)$$

The second integral on the right hand side of (47) can be transformed into a field volume integral by using the divergence theorem.

To do so, we consider the volume between of the whole space *excluding the semiconductor*:

$$\oint_{\Omega_\infty - \Omega_{SC}} div(V \, d\boldsymbol{D}) d\Omega = \int_{\Omega_\infty - \Omega_{SC}} grad(V) \cdot d\boldsymbol{D} \, d\Omega + \int_{\Omega_\infty - \Omega_{SC}} V \, div(d\boldsymbol{D}) d\Omega \qquad (48)$$

leading to:

$$\oint_{\Omega_\infty - \Omega_{SC}} div(V \, d\boldsymbol{D}) d\Omega = - \int_{\Omega_\infty - \Omega_{SC}} \boldsymbol{E} \cdot d\boldsymbol{D} \, d\Omega + \int_{\Omega_\infty - \Omega_{SC}} V \, d\rho \, d\Omega \qquad (49)$$

Setting the potential on the conductor to zero, i.e. $V_C = 0$, and since there is no charge in free space, the second integral in (49) vanishes:

$$\oint_{\Omega_\infty - \Omega_{SC}} div(V \, d\boldsymbol{D}) d\Omega = - \int_{\Omega_\infty - \Omega_{SC}} \boldsymbol{E} \cdot d\boldsymbol{D} \, d\Omega \qquad (50)$$

Since $\Omega_\infty - \Omega_{SC}$ and $\Omega_{SC}$ have opposite surface orientations, using again the divergence theorem and noting that at infinity the displacement vector must vanish, we have:

$$\int_{\Omega_\infty - \Omega_{SC}} div(V \, d\boldsymbol{D}) d\Omega = -\oint_{S_1} V_N (d\boldsymbol{D} \cdot d\boldsymbol{S}) \qquad (51)$$

The incremental work spent by the voltage source is then:

$$\delta W_{vs} = \oint_{S_1} \psi_S (d\boldsymbol{D} \cdot d\boldsymbol{S}) + \int_{\Omega_\infty - \Omega_{SC}} div(V \, d\boldsymbol{D}) d\Omega = \oint_{S_1} \psi_S (d\boldsymbol{D} \cdot d\boldsymbol{S}) - \int_{\Omega_\infty - \Omega_{SC}} \boldsymbol{E} \cdot d\boldsymbol{D} \, d\Omega \qquad (52)$$

The total work spent by the voltage source is obtained by integrating (52) over the displacement vector:

$$W_{vs} = \int_D \oint_{S_1} \psi_S (d\boldsymbol{D} \cdot d\boldsymbol{S}) - \int_D \int_{\Omega_\infty - \Omega_{SC}} \boldsymbol{E} \cdot d\boldsymbol{D} \, d\Omega = \int_{S_1} \int_D \psi_S (d\boldsymbol{D} \cdot d\boldsymbol{S}) - \int_{\Omega_\infty - \Omega_{SC}} \int_D \boldsymbol{E} \cdot d\boldsymbol{D} \, d\Omega \qquad (53)$$

The electric energy supplied by the voltage source is divided in some energy stored in free space and in the body. On the other hand, the electrostatic energy is still given by relation (3). Likewise for the capacitor plate, the difference in these quantities represents an additional free energy for the body:

$$A_{Extra} = -W_{vs} - W_{el} = -\int_{S_I}\int_D \psi_S \left(d\mathbf{D}\cdot d\mathbf{S}\right) - \int_{\Omega_{SC}}\int_D \mathbf{E}\cdot d\mathbf{D}\, d\Omega \tag{54}$$

Relation (54) is the generalization of relation (22) for arbitrary geometries (it is easy to verify that (54) gives the mid-term of (22) in one dimension). Therefore, even though we followed a quite different process, we come to the same conclusion: an additional free energy is stored in matter when this is subjected to an electric field, and only in some special cases this energy disappears.

## VII Conclusion.

Following different theoretical developments, we conclude on the existence of a new electric free energy taking place in semiconductors, and possibly in a variety of materials, which generate errors when dealing with force calculations. Depending on the charge density stored in the body and on the magnitude of the external electric field, such an energy may largely exceed the well-known free energy related to the dipoles polarization process. A generalization to three-dimensional systems is proposed and general rules regarding the need for such a correction are discussed. It comes out that under special situations, as for instance in insulators or when the full depletion approximation holds, this energy vanishes and gives back the commonly accepted definition of electric energy. Besides these fundamental aspects, we anticipate that totally neglecting this energy will generate important errors when evaluating electrostatic forces in micro- and nano-meter scale systems.

# REFERENCES.

1)  Bobbio, Electrodynamics of materials; forces, stresses and energies in solids and fluids (Academic Press, New York, 1999).

2)  S. Bobbio, "The use of energy and co-energy for the evaluation of forces in non-linear, anisotropic dielectric solids", Eur. Phys. J. B **16**, pp. 43-48, 2000.

3)  L.D. Landau, E.M. Lifshitz, Electrodynamique des milieux continus (1990, MIR, Moscou).

4)  Y. Zimmels, "Electromagnetic intensive work and energy variables of finite and composite systems", Journal of magnetism and magnetic materials, vol. 292, p. 433-439, 2005.

5)  John David Jackson, 'Classical Electrodynamics', John–Wiley and Sons, 3$^{rd}$ Edition, 1999 .

6)  Y. Zimmels, "Entropy of electromagnetic polarization", Phys. Rev. E, vol. 65, 036146, 2002.

7)  D.R. Frankl, 'Electrostatic energy in a semiconductor surface space-charge layer', Surface Science, Vol. 9, pp. 73-86, 1968.

8)  S.M. Sze, Physics of semiconductor devices, 2$^{nd}$ edition, John Wiley & Sons.

**FIGURE CAPTION.**

FIG. 1.

Representation of the system of coupled capacitors used in the virtual experiment.

FIG. 2.

Energy representation of the metal-vacuum-semiconductor capacitor structure.

FIG. 3.

Electric field versus surface potential for a p-type doped semiconductor. The surfaces between the curve and the dotted lines represent the integrals defined in relation (22).

FIG. 4a &4b.

Ratio between the extra free energy $A_{Extra}$ and the dipoles energy $U_P$ versus the surface potential (4a) and the charge density in silicon (4b) for different doping concentrations.

FIG. 5a and 5b.

Ratio between the electric force calculated with the standard definition of electric energy ($F_U$) and the Coulomb force ($F_C$) as a function of the charge density and for different doping concentrations. Two values of electrode separation are addressed, namely 100nm (fig. 5a) and 10 nm (fig. 5b).

FIG. 6.

Sketch of a system involving arbitrary conductor and semiconductor bodies.

**Figure 1.**

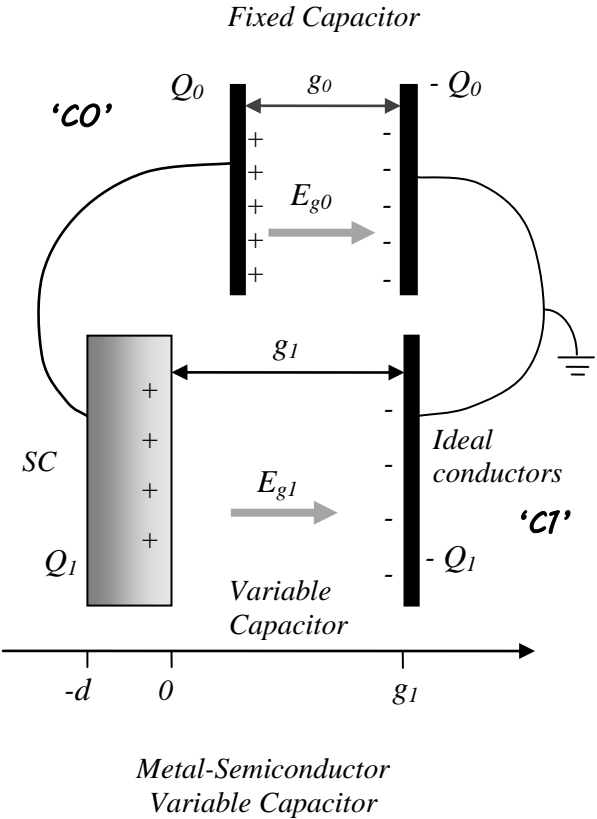

Metal-Semiconductor
Variable Capacitor

**Figure 2.**

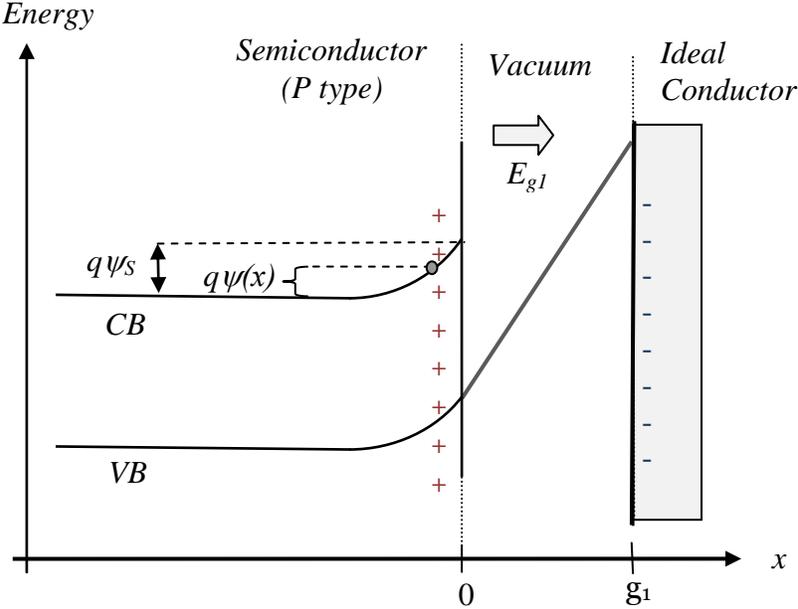

**Figure 3.**

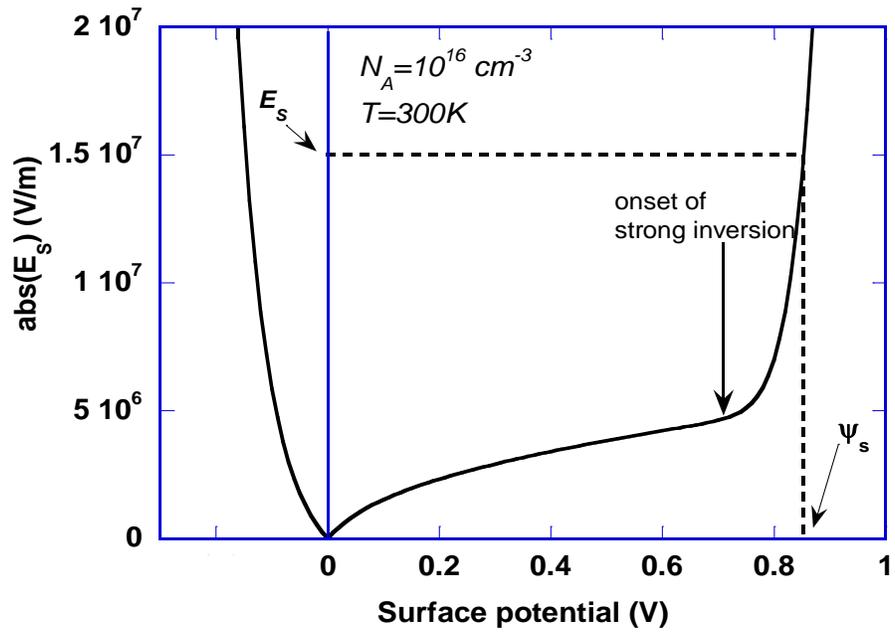

**Figure 4a.**

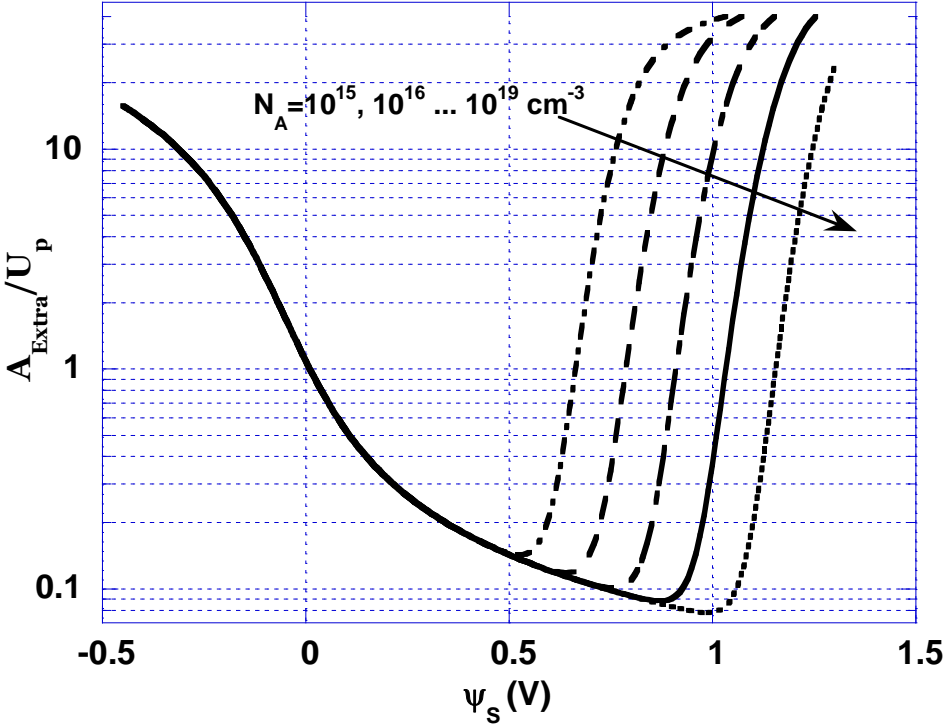

**Figure 4b.**

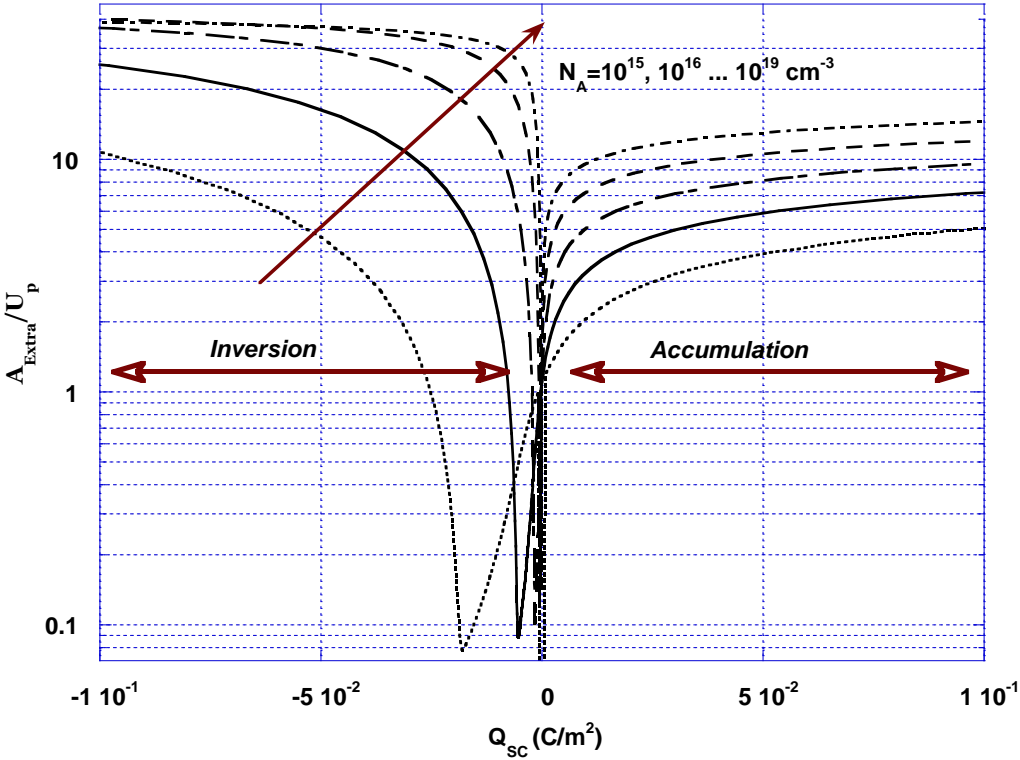

**Figure 5a.**

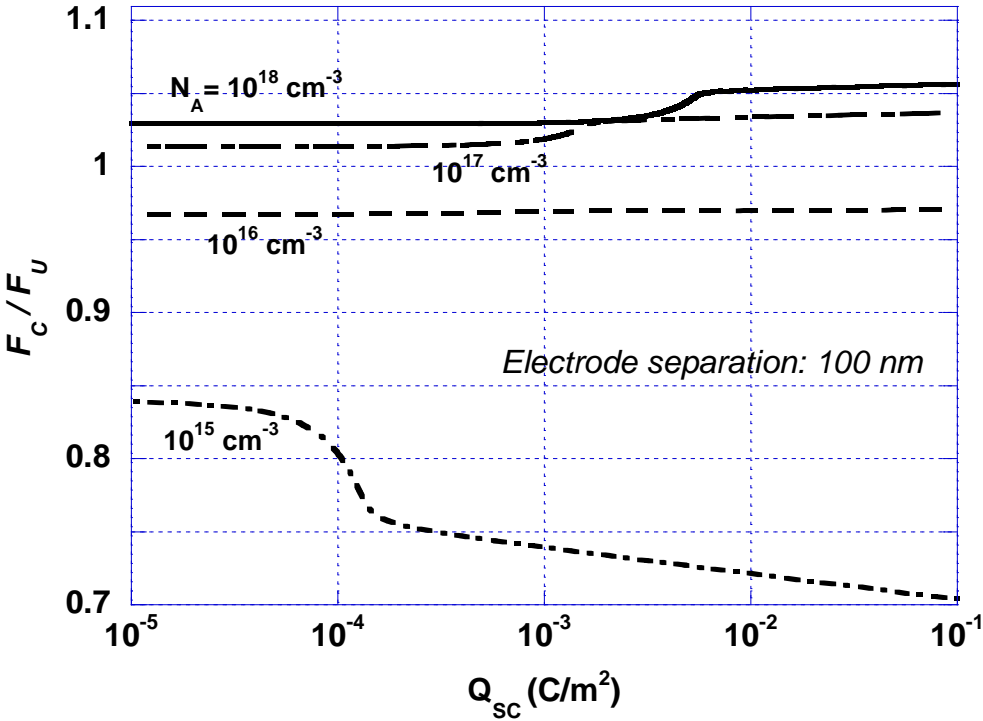

**Figure 5b.**

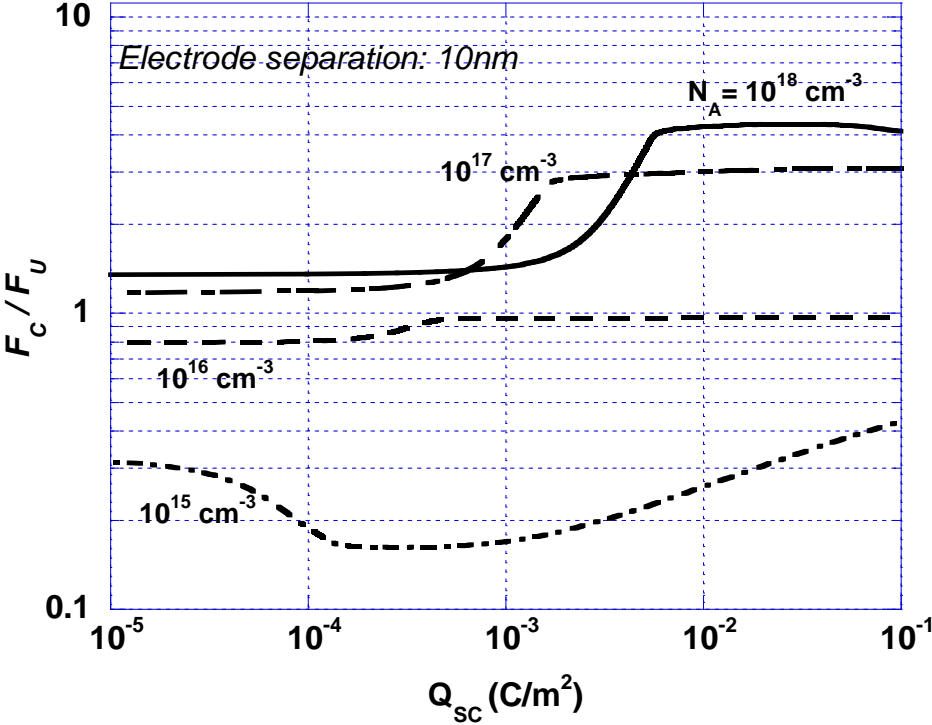

**Figure 6.**

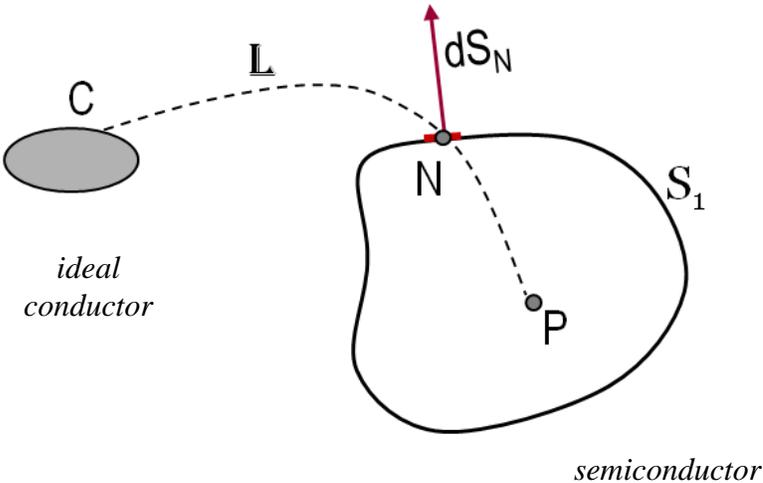